# Toward a Generic Vehicular Cloud Network Architecture: A Case of Virtual Vehicle as a Service

Fekri M. Abduljalil

Department of Computer Science, University of Sana'a, Sana'a, Yemen

*ABSTRACT*

*In recent years, cloud computing has gained more and more popularity. The motivation towards implementing cloud computing in vehicular networks is due to the availability of communication, storage, and computing resources represented by communication, vehicles, roadside units (RSUs), and central servers. These resources can be utilized and provided to vehicles, drivers on the road, travellers, and customers on the internet. Intelligent Transportation System (ITS) applications can utilize vehicular cloud computing to provide efficient real-time services, as well as to improve transportation safety, mobility, and comfort levels for drivers. In this paper, all possible vehicular cloud models are presented. Each vehicular cloud model offers different services. Integrating all vehicular cloud models into one integrated system will provide all services and serve internet users, passengers, and vehicles. Therefore, a generic vehicular cloud model is proposed. After that, a new service called Virtual Vehicle is proposed in vehicular cloud computing. The virtual vehicle is a virtual machine that migrates from one physical vehicle to another. It provides the same services as the physical vehicle according to the consumer's requirements.*

## 1. INTRODUCTION

Cloud computing is a new computing model that provides pools of physical computing resources, known as data centers, to clients as computing services. These services include servers, storage, databases, networking, software, analytics, and more, which can be rented on demand via web browsers. Most cloud computing services fall into three broad categories: Infrastructure as a Service (IaaS) for processing, storage, bandwidth, etc., Platform as a Service (PaaS) for programming languages and operating systems, and Software as a Service (SaaS) for applications such as emergency management, roadway maintenance, electronic payment, and pricing [1].

The VANET is a subclass of the Mobile Ad Hoc Network (MANET) that also has no fixed topology. Vehicles can acquire information and services through Vehicle-to-Vehicle (V2V), Infrastructure-to-Vehicle (I2V), or Roadside-to-Vehicle (R2V) communications. V2V communication is based on Dedicated Short-Range Communications (DSRC) technology, while I2V communication is based on GPRS/3G, Wi-Fi, or WiMAX. The emergence of new wireless systems, Intelligent Transportation System (ITS) applications and services, and ITS technologies has raised novel research questions on both safety and non-safety-related applications [2].

Vehicular Cloud Computing (MCC) is a new paradigm that aims to provide on-demand services as a utility anytime, from anywhere, and at low cost through a pay-as-you-go model. The motivation for cloud computing in vehicular networks is the communication, storage, and computing resources available in the vehicular network, which include communication devices, vehicles, roadside units (RSUs), and central servers. These resources can be utilized and provided to vehicles, drivers on the road, travelers, and customers on the internet [3]. Research studies have stated that the vehicular cloud can provide many different services, including "Network-as-a-Service" (NaaS), "Collaboration-as-a-Service," "Storage-as-a-Service" (StaaS), "Sensing-as-a-Service" (SeaaS), content sharing, and traffic management [3].

There are many service consumers and application managers interested in renting vehicular cloud services and resources, as well as obtaining information collected by vehicles. Many vehicular cloud applications and services depend on vehicle location, such as Pic on Wheel [4], Video Capture [5], vehicle as witness [6], vehicle tracking [7], and others. The vehicular network topology is dynamic, with vehicle nodes parking, moving, and changing location according to driver interest. Vehicle location has a great effect on services and applications provided to service consumers. The challenge is how to make vehicle services and applications run according to service consumer location requirements. Vehicular cloud computing faces many challenges, such as the production of a generic network architectural framework and the provision of many different services based on mobile computing. This research paper studied these challenges and it presents a solution for these challenges.

In this paper, we propose a new vehicular cloud network architecture. We also propose a new service called Virtual Vehicle, which enables service consumers to control vehicle movement and obtain information from services and applications specific to a location. The Virtual Vehicle is a virtual machine that runs all applications requested by service consumers. The virtual machine migrates from one physical vehicle to another in order to satisfy service consumer location requirements.

The rest of this paper is organized as follows: Section 2 presents related works. Section 3 describes the vehicular cloud system model. Section 4 presents the concept of the Virtual Vehicle. Finally, Section 5 concludes the paper with remarks for future directions.

## 2. RELATED WORKS

In [8], the paper proposes the integration of Vehicular Ad Hoc NETwork (VANET) with cloud computing in order to establish a novel concept named Vehicular Cloud Computing (VCC). The paper introduces a comprehensive taxonomy of VANET-based cloud computing, which represents a pioneering initiative in defining VANET Cloud architecture. In particular, the VANET clouds are categorized into three distinct architectural frameworks, namely Vehicular Clouds (VC), Vehicles using Clouds (VuC), and Hybrid Vehicular Clouds (HVC). Additionally, the paper outlines the unique security and privacy issues as well as the research challenges that arise in the context of VANET clouds.

In the paper [9], a multi-layered context-aware architecture for cloud-assisted vehicular cyber-physical systems is presented. The article introduces two essential service components, namely vehicular social networks and context-aware vehicular security. Moreover, it proposes an application scenario for context-aware dynamic parking services using the cloud-assisted architecture and logic flow. The paper delves into the challenges and possible solutions for context-aware safety hazard prediction, context-aware dynamic vehicle routing, and context-aware vehicular clouds. The methods employed in this paper are primarily conceptual and theoretical, as the authors put forth a new architecture and discuss the challenges and solutions based on their expertise in the field. The paper does not involve any empirical or experimental methods.

In the article by [10], the notion of vehicle cloud computing is introduced, with a focus on its implementation in vehicular networks. The authors explore recent research in the field and propose a taxonomy for vehicle cloud architecture based on architecture and service types, aimed at identifying and addressing any remaining shortcomings. The paper elucidates the use of vehicle Ad Hoc Network (VANET) as a type of wireless ad hoc network that facilitates connection between vehicles and Road Side Units (RSUs) along the road. Additionally, it emphasizes the central objective of VANET, which is to provide seamless connectivity to mobile users while on the road and to enable efficient Vehicle-to-Vehicle (V2V) communication, thereby enhancing the Intelligent Transportation System (ITS). Finally, the authors describe Cloud Computing (CC) as a novel paradigm in the computing realm that offers ubiquitous, applicable, and on-demand network access to an extensive array of shared computing resources, including networks, servers, storage, applications, and services.

The present study [11] introduces a novel classification system for Vehicular Cloud Architectures (VCAs), which is grounded in two distinct delineations: service delivery paradigm and Quality of Service (QoS) awareness. The study aims to investigate security-related concerns that must be taken into account in VCAs. The research endeavors to identify the diverse hurdles that necessitate resolution in VCAs. The study furnishes a comparative appraisal of the scrutinized architectures. No explicit techniques employed in the investigation are described in the study.

The article [12] delves into the utilization of onboard resources for transportation systems, as well as the advancements in management technology for Cloud computing resources. This paper presents the most recent approaches and solutions for Vehicular Clouds, including applications, services, and traffic models that can enable Vehicular Cloud in a more dynamic environment. The article examines numerous applications and services that have relevance in the transportation system, benefiting management, drivers, passengers, and pedestrians. The paper analyzes existing traffic models and determines that Vehicular Cloud computing is technologically feasible not only in static environments such as parking lots or garages where vehicles are stationary, but also in dynamic scenarios such as highways or streets where vehicles are in motion. The paper abstracts current issues and classifies solutions as a matter of leveraging underutilized vehicular resources, such as network connectivity, computational power, storage, and sensing capability, which can be applied to the Vehicular Cloud. Although no specific research methods are presented in this paper, it is focused on discussing the concept and solutions related to Vehicular Clouds.

In paper [13], an extensive survey of current research on vehicular cloud computing is provided. It highlights several key issues of vehicular cloud such as architecture, inherent features, service taxonomy, and potential applications. The paper contributes to the field of vehicular cloud computing by providing a comprehensive overview of the current state of research in this area. It also identifies several research challenges and opportunities for future work. Overall, the paper provides a valuable resource for researchers and practitioners interested in vehicular cloud computing. The paper does not involve any experimental or empirical methods.

The present study [14] scrutinizes diverse vehicle architectures with the aim of efficiently accommodating the escalating number of software-reliant vehicle services. It propounds the notion of a software-defined vehicle and arranges essential connected vehicle applications based on their performance prerequisites. The study incorporates a case study on over-the-air updates and scrutinizes four categories of connected vehicle architectures: centralized, decentralized, publish/subscribe, and broadcast. It deliberates the benefits and drawbacks of each architectural variety and culminates by presenting an edge-based architecture for connected vehicles christened EdgeArC.

The paper presented in reference [15] highlights various research issues and challenges that must be tackled to ensure the efficient functioning of Vehicular Cloud Computing Networks (VCCN). The paper advocates for the implementation of a dynamic offloading and resource provisioning plan, a proficient operational model for Vehi-Cloud, a framework for system model and architecture, and methods for synchronization of Vehicular Cloud and aggregation of data to address these challenges. The paper suggests that there is a need for further research in these areas to enable dynamic access to hybrid and community clouds.

In the work of paper [16], a proposal is put forth for a four-layered architecture for cloud computing in VANETs that encompasses perception, coordination, artificial intelligence, and smart application layers. The proposal delves into the three network components that comprise cloud computing in VANETs, namely, vehicle, connection, and computation, and elucidates their cooperative roles. Furthermore, a taxonomy for cloud computing in VANETs is presented, which takes into account the major themes of research in the field, such as the design of architecture, data dissemination, security, and applications. Recent advancements and challenges in vehicular cloud computing (VCC) and vehicle using cloud (VuC) are reviewed in the paper, along with a summary of related literature on the design of architecture, protocols, techniques, implementation, and remarks. Finally, the potential future challenges and opportunities in cloud computing in VANETs are discussed.

The paper in [17] provides an overview of the motivation behind vehicular cloud computing. It identifies challenges related to the design of vehicular cloud computing. It highlights the features of existing vehicular cloud architectures and providing a taxonomy of vehicular cloud. It discusses open research directions in the field of vehicular cloud computing.

In [18], a new model for the vehicular cloud called VANET-Cloud is proposed to improve traffic safety and provide computational services to road users. This model extends the traditional cloud infrastructure, which consists mostly of stationary nodes, to the edge of vehicles.

In [19], an architecture for the vehicular cloud is presented. It describes a static vehicular cloud that can be formed in any parking lot, airport, or shopping mall. The architecture also explains the functions of data storage and effective retrieval of data from the data center in the vehicular cloud.

In [1], a novel concept called the Autonomous Vehicular Cloud (AVC) is presented. AVCs utilize advancements in mobility, powerful embedded in-vehicle resources, ubiquitous sensing, and cloud computing to enhance the safety, security, and economic vitality of modern society.
A new VANET network planning is introduced in [9][20]. It consists of two paradigms: Vehicular Cloud Computing and Information Centric Networking. This research proposes a future vehicular networking system called Vehicular Cloud Networking. The design principles of Vehicular Cloud Networking are built on top of Vehicular Cloud Computing and Information Centric Networking. Vehicular Cloud Computing enables vehicles to discover and share their resources to create a vehicular cloud with value-added services, while Information Centric Networking enables efficient distribution of cloud contents among vehicles.

In paper [21], a new concept of a data center in an airport is proposed. The idea involves utilizing cars in the long-term parking lot of a typical international airport to build a vehicular cloud.
In paper [22], the vehicular cloud is used as virtual edge servers for efficient connections between cars and backend infrastructure for future Intelligent Transportation System (ITS). This research introduces the concept of vehicular micro clouds, which are clusters of cars that collect and transfer data to backend infrastructure. An enhancement and study for this idea is presented in [23].

## 3. VEHICULAR CLOUD NETWORKS MODEL PROPOSALS

In this section, we present the different proposed models for vehicular network architecture, as well as our generic model for the vehicular cloud.

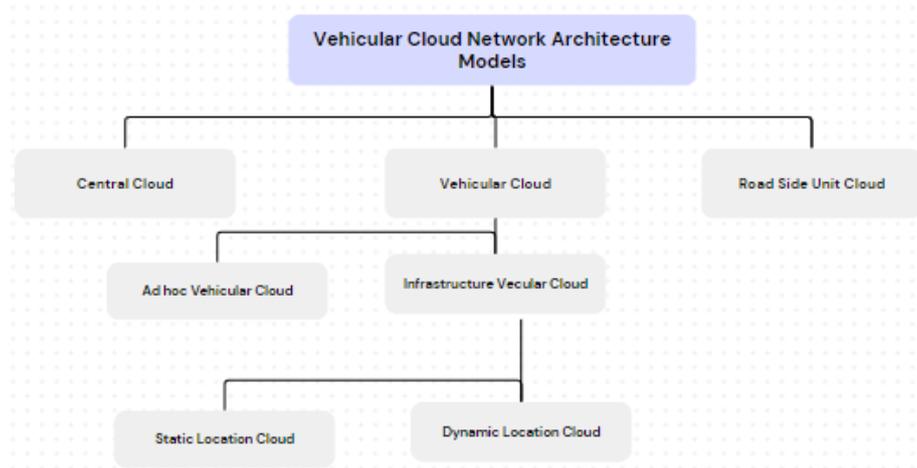

Figure 1. Types of vehicular cloud models.

### 3.1. Vehicular Cloud Network Architecture Models

In this section, we present the different possible architecture for cloud in Vehicular Cloud Computing. According to previous studies, the vehicular cloud can be divided into five types as shown in figure 1.

### 3.1.1. Ad Hoc Vehicular Cloud (Mobile Vehicular Cloud)

Figure 2 shows the architecture model for this cloud. It consists of a group of vehicles, each of which has resources such as computing and communication capabilities that can offer services. One of the vehicles becomes the service provider, called the broker [24], and other vehicles can join the cloud according to a contract between the service provider and the vehicle drivers. The service consumer can be any vehicle in the location of the cloud, and passengers in the vehicles can also be service consumers. The services provided by this cloud include software as a service, road information, and others. Users on the internet can access this cloud if mobile IP is supported and access points are available. The Vehicular Cloud is a number of mobile servers with dynamic topology.

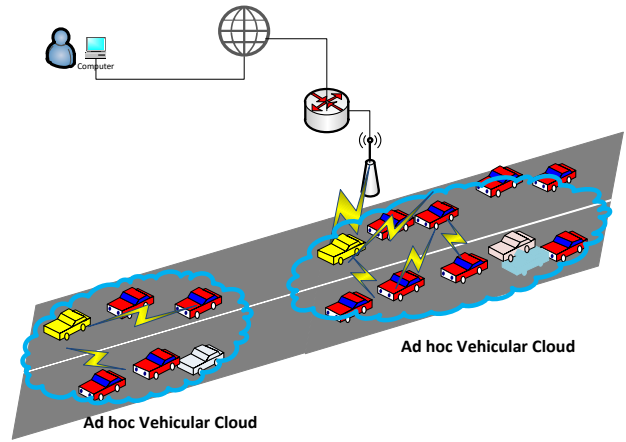

Figure 2. Ad hoc vehicular cloud model

### 3.1.2. Stationary Infrastructure Vehicular Cloud (Temporary Stationary Vehicular Cloud)

Figure 3 shows the architecture model for this cloud. It consists of a group of vehicles that are parked in locations such as parking lots or shopping malls and is managed by a server owned by the service provider, which may be a company or organization [21]. The cloud can be accessed by service consumers in the location of the cloud, which may belong to a university, airport, shopping mall, or other institution. This type of cloud can be implemented as a private cloud. The cloud services offered can be accessed by users in the location of the cloud or by users on the internet. The vehicles can be rented by the service provider according to their needs for resources [24].

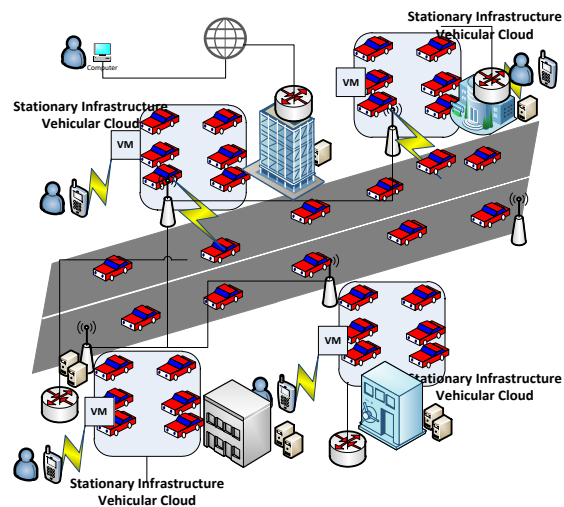

Figure 3. Stationary Infrastructure Vehicular Cloud

### 3.1.3. Dynamic Infrastructure Vehicular Cloud

Figure 4 shows the architecture model for this type of vehicular cloud. It consists of mobile vehicles serving as physical servers and a stationary server at a roadside unit [25]. The stationary server is used to manage the cloud and it is the service provider server. This cloud contains one or more access points and may contain more than one roadside unit. The cloud offers its services to vehicles and users on the internet [24].

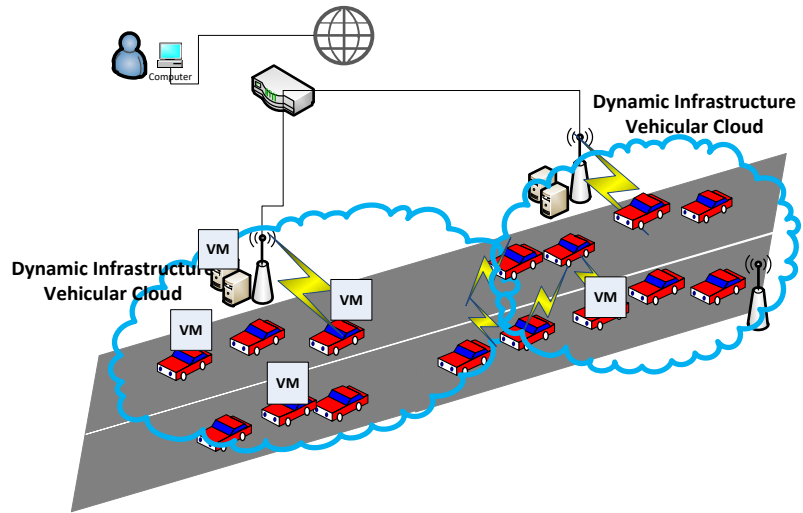

Figure 4. Dynamic Infrastructure Vehicular Cloud

### 3.1.4. Roadside Cloud

We include this cloud in the vehicular cloud because its objective is to offer services to vehicles or passengers in vehicles. Figure 5 shows the architecture model for this cloud. The RSU Cloud is a group of physical servers and wireless communication towers located in close proximity to the roads. It serves the vehicular networks and includes more than one access point and more than one roadside unit. The service provider manages the servers in one or more roadside units, and the cloud offers services to vehicles or passengers in vehicles [26].

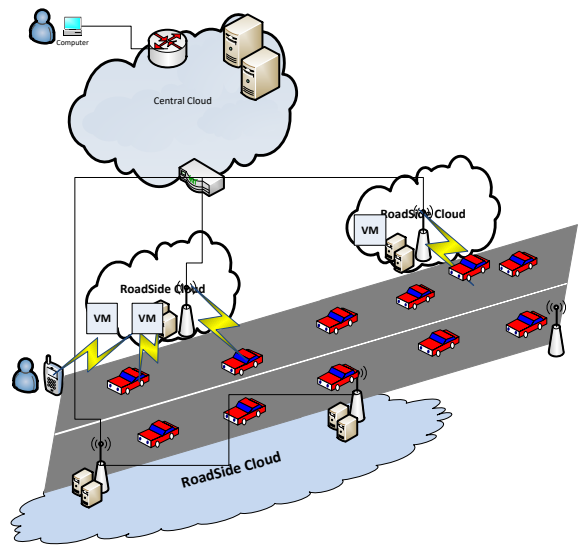

Figure 5. Road Side Cloud

### 3.1.5. Central Vehicular Cloud

The central cloud is a group of stationary physical servers located in data centers on the internet. The servers are housed in remote data centers, and the service provider offers basic services to users or vehicles. The service provider manages servers that can be placed in any data center on the internet, and it offers services according to consumer requests and depending on the service. In this architecture, the cloud has a group of stationary dedicated servers and can add mobile physical servers during searches for resources. First, the vehicle registers with the central cloud service provider. Then, when a service provider receives a request for a service from a vehicle, it searches the registered vehicles and uses them [18].

### 3.2. Generic Model for Vehicular Cloud

As we have seen in the previous section, there are five models for building clouds for vehicular networks. We propose combining all five models into one architecture model, as shown in Figure 6. The main parts of the cloud system are the service provider, service consumer, and resources. The integrated system serves internet users, passengers, and vehicles. In this system, there are two types of physical servers: traditional stationary servers in the central cloud and roadside units, and mobile physical servers in the vehicles.

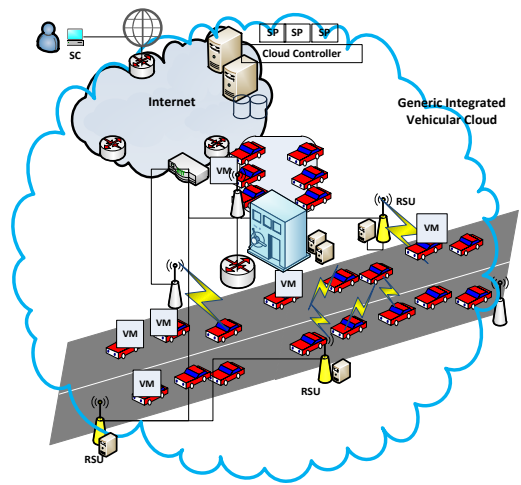

Figure 6. Generic VC Network Architecture Model

Managing each architecture model is easier than the integrated model. We suggest a level of management for each level related to the number of managed resources (vehicles). The service consumer can see two types of clouds: services offered from physical servers in data centers and services offered from vehicles. The service provider has three types of services. The proposed generic model architecture shown in Figure 6 consists of the following entities:

**Service Provider Center Manager (SPCM):** The SPCM manages all service providers in the cloud and serves as the representative of service providers, acting as a medium between vehicles and service providers. It is responsible for all of the cloud's main operations, such as searching for resources, provisioning resources, and virtual machine migration.

**Service Provider (SP):** It is responsible for allocating and managing services, such as advertising services, making contracts for services, validating proofs of work done by vehicles, and tracking vehicles and their information.

**Service Consumer (SC):** It is a user on the web requesting cloud services or requesting a virtual vehicle.

**Road Side Unit (RSU):** The RSUs are the stationary nodes of vehicular cloud. They are located along the roads and connected by a network so they serve as gateway to the central cloud vehicular cloud. RSU is equipped with computational resources and storage facilities, and can be utilized.

**Physical vehicle (OBU):** Physical vehicle has An OBU system installed on it. And it has computation, sensing, communication capabilities and storage. each connected vehicle represents a mobile physical host in a data center.

**Virtual Vehicle (VV):** it is a virtual machine that can be moved or driven according to the requirements of the user in the road, depending on the movement of physical vehicles to collect information or monitor the road and other services. The virtual machine executes different application as services. The service consumer accesses virtual machine and application through SP web portal.

We have three scenarios for Service Provider Manager (SPM) placement. The SPM servers can be placed in any data center on the internet, offering services according to consumer requests and depending on the service. The SPM servers can also be placed in any RSU physical server, managing all servers in its RSU or other RSUs and vehicular nodes. Finally, the SPM servers can be placed in any vehicle, managing all servers in the vehicular cloud.

In the proposed model, there are two types of physical hosts in the cloud. The first type is the stationary host, located in the data center of the central cloud and the servers in the RSU cloud. The second type is the mobile physical host, which can be any vehicle in the vehicular cloud.

The system installed in vehicles supports virtualization, with a Virtual Machine Manager (hypervisor) used to manage virtual machines. It has a network connection with other mobile vehicles (V2V) and RSUs (V2I). The service consumer accesses vehicular cloud services through a web-based interface implemented using web services, portals, or REST API. The service consumer can directly access a vehicle to get service, or it can access a data center that includes data collected from vehicles and offered as services. The type of service guides the cloud controller to select the resource location. IP mobility management is handled using Mobile IPv6. The service center server (cloud manager) must be running to offer service to the service consumer. Vehicles interested in offering services can join the cloud and can also leave the cloud. If a joined vehicle does not respond, the cloud controller deactivates it. The joined vehicle updates its information periodically to the cloud controller, which maintains the vehicle's information and uses it to search for resources when a service consumer requests service. Vehicles join the cloud using a registration request, sending a list of resources and services that can be offered. The cloud controller is a server in either a data center, an RSU, or a vehicle. The cloud controller server address can be offered manually to vehicles if the cloud controller is in a data center on the internet. If the cloud controller server is in an RSU or a mobile vehicle, the vehicle discovers it using a discovery algorithm.

If a vehicle wants to leave the cloud, it sends a leave request to the cloud controller. If the vehicle receives a leave response, it migrates all virtual machines to the destination vehicle. Otherwise, the vehicle migrates all virtual machines to the RSU. The response includes the destination vehicle for virtual machine migration.

Two main issues are considered in virtual machine migration, which are virtual machine migration decision and selection of the candidate vehicle. The virtual machine migration decision can be made by either the cloud controller or the vehicle. There are many reasons for migration, such as a vehicle leaving the cloud, a vehicle leaving the service zone, or workload sharing between vehicles. The candidate vehicle selection can also be made by either the cloud controller or the vehicle. The selection criteria can include random selection, vehicle with low workload, remaining time in zone, or criteria based on the service offered.

## 4. THE PROPOSED VIRTUAL VEHICLE AS SERVICE

The detail operations of the proposed service are presented in this section.

### 4.1. Vehicle Registration with SPCM

Each vehicle registers itself with the service provider cloud manager, indicating that it is ready to offer its computation and storage resources.

### 4.2. Vehicle Location Update

The vehicle updates its location, direction, and speed at regular intervals, such as every 10 minutes. The cloud manager maintains this information about every vehicle in the cloud.

### 4.3. Request Virtual Vehicle

The user can request the service provider to provide a virtual vehicle that can be controlled and driven on a specified road. The user request message includes the source location, destination location, speed, and direction. The service provider searches for available vehicles and selects a physical vehicle according to the user request. If the list of vehicles satisfies the user request, a vehicle can be selected randomly. The service provider sends a message to the selected vehicle to create a virtual machine with the user-requested services and applications. This virtual machine will run all applications and services requested by the user, such as video capture, information about roads, and messages received from nearby vehicles. The service provider keeps information about each virtual machine created.

### 4.4. Driving Information Update

The user must continuously communicate with the virtual vehicle (VM) to guide and control its movements. As mentioned in the previous section, the user request message includes the source location, destination location, speed, and direction. The user provides information such as the destination location, speed, and direction to the service provider. The service consumer may change the driving information, so the driving information should be updated according to the user request. The service consumer sends the new information to the service provider. The user communicates directly with the virtual vehicle to get real-time road information.

### 4.5. Virtual Machine Migration

Like traditional cloud computing, virtual machine migration (VMM) is used in vehicular cloud computing to solve different problems and provide services such as load balancing and power saving. However, in vehicular cloud computing, the connected vehicles represent physical servers that move at varying speeds. As a result, the vehicular network topology changes more rapidly, and physical hosts have limited computing and storage capacity compared to a central cloud.

Therefore, VMM is an important function in vehicular cloud computing. The virtual machine migration steps proposed in our system are as follows:

### 4.5.1. Migration Decision

As mentioned in the previous section, the vehicle periodically updates its information to the service provider cloud manager (SPCM), which can be accessed by the service provider (SP). The SP makes a migration decision when the physical vehicle parameters, such as speed and direction, are not suitable for the user request. The driving parameters are the set of variables given to the service provider and used for virtual machine (VM) migration from one physical vehicle to another. The user provides information such as the location, direction, and speed to the service provider. The physical vehicle's real-time information, such as location, direction, and speed, is sent periodically to the SPCM and can be accessed by the SP. The direction can be modelled as eight values. The speed should be modelled as slow (0-20 km/h), medium (20-50 km/h), and fast (50-120 km/h). The physical vehicle's location can be modelled as same (difference is less than 8 meters), near (difference is between 8 and 30 meters), or far (difference is more than 30 meters). If any of the previous information does not match the requested virtual vehicle information, the SP searches for a candidate vehicle. The requested location of the virtual vehicle will be the location of the physical vehicle.

### 4.5.2. Vehicle Candidate Selection

The service provider (SP) searches its database for candidate vehicles that satisfy the user's information. A candidate vehicle is any vehicle with a location nearest to the requested location and has the same modeled direction and speed. The SP requests virtual machine (VM) migration to the best vehicle candidate. In the worst case, the VM migrates to the roadside unit (RSU) and then to the best available vehicle candidates. The user provides input such as the start location, speed, direction, and destination location. The vehicle updates include location, speed, and direction. For each user request, if the virtual machine is not created, then the service provider searches the vehicle list, creates a zone, selects a vehicle host, and creates the VM in the vehicle. 'U' is a variable representing the accepted distance from the host vehicle to be in the zone. For each vehicle update, if the distance from the vehicle host is 'U', then the vehicle is added to the zone. For each update from the host vehicle, if the direction or speed changes, then the VM is migrated to the new nearest host in the zone with the same user information

______________________________________________________
**Algorithm 1**: Vehicle Candidate Selection                .
**Input:** current RSU, current location, speed, Direction, U
**Output:** new Vehicle ID
  For each vehicle in current RSU{
       Get vehicle
     If (distance <U)
     Add vehicle to zone
  }
   4:Find vehicles in zone with minimum workload and same speed, Direction
 If found then
       Return vehicle with minimum workload
       Start VM migration to new vehicle
  Else
       Start VM migration to current RSU.
        Goto 4
  End if
______________________________________________________

**4.5.3. Resources Reservation**

The service provider cloud manager (SPCM) sends a message to the candidate vehicle for resources reservation and initiates a container virtual machine (VM) on the candidate vehicle. The VMs are copied in successive rounds. An acknowledgment message of successful migration should be sent to the SPCM, which should be aware of any unsuccessful migration. The candidate vehicle should inform the old vehicle and the service provider that it has successfully received the migrated VMs. The old vehicle acknowledges this message as a commitment of the migration transaction. The old vehicle may now discard the original VMs, and the candidate vehicle becomes the primary host. The migrated VMs on the candidate vehicle are now activated. Post-migration code runs to reattach device drivers to the new vehicle and advertise moved IP addresses if required.

## 5. PERFORMANCE ANALYSIS

To verify the efficiency of the proposed system, the researchers in [27] conducted a study to confirm that virtual machine migration could maintain the service in the cloud-based vehicular network, ensuring that the network is continuous and the service is not interrupted during user operations.

Two algorithms, the selection algorithm and the migration process, were implemented using simulation. The simulation program was written in Python. Initially, the virtual machine was created and configured, and then a service request was sent. The remaining time for vehicles in the zone was calculated. The simulation was used to evaluate the performance of each proposed algorithm based on several metrics by proposing inputs and analysing the results.

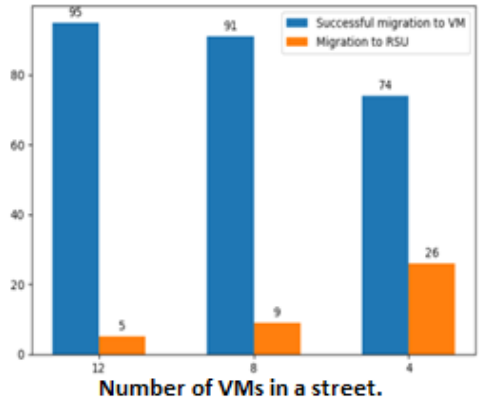

Figure 7. The percentage of successful migration to VM or RSU.

Figure 7 displays the percentage of successful virtual machine (VM) migrations from one vehicle to another vehicle. The graph indicates that if the number of vehicles is large, the successful migration rate to another vehicle will be high. Conversely, if the number of vehicles in the street is small, the number of successful VM migrations to the roadside unit (RSU) will increase.

There are two recommended ways to implement and evaluate the proposed virtual vehicle as a service. First, it is recommended to implement the proposed service in the network simulator (NS 3.27) [28], which comprises the modules of the vehicular network interface, wired networks, and the cellular network. The proposed system should be implemented as a custom application in application layer, with the virtual machine modeled as a file with information. The application implements a selection algorithm and migration alogrithm. TCP socket library can be used to send and receive virtual machines. Moreover, the Simulator for Urban MOBIL 0.25.0 (SUMO) [29] can be used for mobility abstraction using a real trace.

The second way is to implement the proposed service in a testbed as a private cloud using wireless LAN 802.11, Microsoft Hyper-V, and Service Center Virtual Machine Manager (SCVMM) [30]. To create a vehicular cloud, a private cloud should be installed. Then, the virtual vehicle service can be implemented as web services. A laptop connected to the private cloud can be represented as a physical vehicle. The migration process can be handled using SCVMM, and Hyper-V should be implemented in every laptop. Finally, different scenarios can be studied.

## 6. CONCLUSIONS

Vehicular cloud evolves with new services, and this article examined the very details behind vehicular cloud Network Architecture. We looked into vehicular cloud papers and observed different vehicular cloud architecture models, which are used by the different research papers. the different vehicular cloud architectures are studied. Then, a new generic vehicular cloud network architecture is presented which integrate the different vehicular network architecture in one generic model. After that a new cloud service is proposed to be used with the generic vehicular cloud architecture. It is called virtual vehicle as a service. The design details of the proposed service including registration, candidate selection, and migration process are discussed. Finally, different implementation methods to evaluate the proposed system are mentioned as points for further works.

## AUTHOR

**FEKRI M. ABDULJALIL** (f.abduljalil@su.edu.ye) is an associate professor in the Dept. of Computer Science, Faculty of Education, Arts, and Science - Khawlan, University of Sana'a. He received his Ph.D. degree in computer science from Pune University in June 2008. His research interests include IP Mobility Management, Mobile Ad hoc Routing Protocols, Vehicular Networks, Intelligent Transportation System, Cloud Computing, networks management, Vehicular cloud Computing, Network Security, and Internet of Things. He is an IEEE senior member.

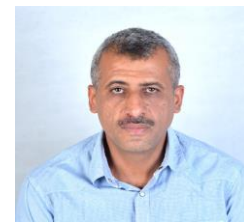